\documentclass[10pt, letterpaper]{article}
\usepackage{graphicx}
\usepackage{amsmath}

\begin{document}

\author{V. V. Prosentsov\thanks{%
e-mail: vitaly.prosentsov@gmail.com} \\
Laan door deVeste 12, 5708 ZZ, Helmond, The Netherlands}
\title{Wave scattering by objects made of small particles with oscillating permittivity}
\maketitle

\begin{abstract}
Rapid advancements in the micro and nano-technology create unlimited opportunities for design of novel optical materials and their applications. Recently, the possibility of the fast refractive index modulation was
demonstrated in semiconductors. In this paper we study the wave scattering by small dispersionless particles with periodically varying refractive index in scalar case by using the local perturbation method. The used formalism allows us to study theoretically and numerically the scattering by objects made of small particles of arbitrary shape and with oscillating refractive index.

In this work, the field scattered by the cluster of the particles and its resonance frequencies are calculated theoretically. In addition, the results of the numerical modeling of the scattering by single cube and by cluster of cubes with oscillating permittivity are presented. It was shown that the scattered fields and their resonance frequencies are significantly affected by the oscillating permittivity: existing resonances are shifting, new scattering resonances are emerging, and deeps in the scattering spectrum are appearing.
\end{abstract}

\section{Introduction}
Light scattering is very broad and classical topic of Optics, and it is extensively discussed in the
literature \cite{Born}, \cite{Jackson}. The scattering by small particles is one of the subtopics of the light
scattering, and it has many applications in scatterometry, optics of meta materials, and
contamination detection (see, for example \cite{Hulst}, and references wherein).

Till recently, the small particles were studied as ones having constant refractive index, while their shapes could be varied broadly \cite{Hulst}-\cite{Chaumet}. In addition, the scattering by the moving particles was well established
topic long ago, and it became fruitful branch of the Optics \cite{Pecora}. The next next logical step would
be to study the scattering by particles with dynamically varied refractive index. Indeed, the number of
such investigations is surged recently (see, for example \cite{Asadchy}-\cite{Ptitcyn}, and references wherein).

In works \cite{Asadchy}-\cite{Ptitcyn}, the scattering by the sphere with time varying refractive index was studied. In the work \cite{Martijn} it was shown that the refractive index of the scatterers
can be changed in time (by illuminating them with THz radiation, for example) and as result, the permittivity of the
particles can be changed with the predefined frequency. It would be interesting to study the wave scattering by the system of arbitrary shaped particles with the time dependent refractive index, and to compare their properties with stationary case when refractive index is constant in time.

The local perturbation method (LPM) is well suited for the theoretical investigation and
numerical modeling of the scattering by arbitrary objects made of small particles (particles which characteristic size is smaller than incident wavelength).  Previously, the LPM was used to study the light scattering by small particles with constant refractive index \cite{Markel}-\cite{Draine}, and the light scattering by the moving particles \cite{DynScatt}.
In this work, I will use the LPM to study the wave scattering by the objects made of the small particles with the oscillating permittivities.

In this paper I study analytically and numerically the wave scattering by the cluster of small dispersionless particles which refractive indexes oscillate with fixed frequencies. Each small particle may have its own permittivity and oscillating frequency with which the permittivity of the particle changes.
By using the local perturbation method (LPM), the explicit expression for the field scattered by the cluster of the small particles with oscillating refractive indexes is calculated in scalar approximation. The analytical expressions for the resonance frequencies of the particles are found when the oscillating frequency is much smaller compared to the optical one, and refractive index variation is small. The resonance width is also calculated.
Moreover, the results of the numerical modeling are presented to show the significance of the oscillating permittivity: our results suggest that the oscillating refractive index may work as terminator and as amplifier of the scattering.

\section{The theoretical formalism}
In this section we study the light scattering by cluster of the small dispersionless particles with oscillating refractive index in scalar approximation. The scalar approximation allows to show the main features of the scattering
 process avoiding at the same time the complexity of the vector case.

Since any scattering object can be considered as one made of small particles, our approach, in principle, can be used for study of scattering by objects with arbitrary shapes and sizes.

 The equation describing the electric field $E$ propagating in the host medium filled with $N$
particles has the form
\begin{equation}
\bigtriangleup E(\mathbf{r},t)-\frac{\partial^2}{c^2\partial t^2}\varepsilon(\mathbf{r},t)%
E(\mathbf{r},t)=\frac{4\pi}{c^2}\frac{\partial}{\partial t}j(\mathbf{r},t). %
 \label{osc1}
\end{equation}
Here $\Delta$ is the Laplacian operator, $\mathbf{r}$ is the radius vector of the observer, and
$t$ is time at the observer's location, $c$ is the light velocity in vacuum, and $j$ is the field source.
The permittivity of the medium filled with the particles is denoted as $\varepsilon(\mathbf{r},t)$, and it can be presented in the form
\begin{equation}
\varepsilon(\mathbf{r},t)=\varepsilon _{h}+%
\sum_{n=1}^{N}f_{n}(\mathbf{r}-\mathbf{r}_{n})(\varepsilon _{n}(t)-\varepsilon _{h}), %
\label{osc2}
\end{equation}
where $\varepsilon _{h}$ is the relative (in respect to vacuum) permittivity of
the host medium, and the function $f_{n}(\mathbf{r}-\mathbf{r}_{n})$ describes the shape of the $n$-th particle
(with the characteristic size $L_{n}$) as
\begin{equation}
f_{n}(\mathbf{r}-\mathbf{r}_{n}) = \left\{ \begin{array}{cc}%
1\text{,} & \text{inside particle } \\ 0\text{,} & \text{outside particle.}%
\end{array}  \right.  %
\label{osc3}
\end{equation}
Here $\mathbf{r}$ and $\mathbf{r}_{n}$ are the radius vectors of the observer and the $n$-th particle respectively. The permittivity of the $n$-th particle $\varepsilon_{n}$ is set as oscillating function with offset $\varepsilon _{n}^{0}$,
the variation amplitude $\delta\varepsilon _{n}$, and zero phase as
\begin{equation}
\varepsilon_{n}(t) = \varepsilon _{n}^{0} + \delta\varepsilon _{n} \sin^2(\Omega_{n} t/2), %
\label{osc4}
\end{equation}
where $\Omega_{n}/2$ is the angular frequency of oscillations.

Since we consider the scattering by small particles (or scattering by objects made of small particles), the local perturbation method (LPM) is suitable tool to investigate this kind of problem. By applying the LPM approach, we use the following relation between the field $E(\mathbf{r},t)$ and the field $E(\mathbf{r}_{n},t)$ inside the small particle
\begin{equation}
\varepsilon(\mathbf{r},t)E(\mathbf{r},t) \approx \varepsilon(\mathbf{r},t)E(\mathbf{r}_{n},t)
\label{osc5}
\end{equation}
for the field $E(\mathbf{r},t)$ in the Eq. (\ref{osc1}). After this, we integrate the modified Eq. (\ref{osc1})
over time $t$, and make use of the Fourier transform in the frequency domain. Finally, we obtain the following
wave equation for the field $\widetilde{E}(\mathbf{r}, \omega) $ in the space-frequency domain
\begin{eqnarray}
(\bigtriangleup + k^2) \widetilde{E}(\mathbf{r}, \omega) + \frac{k^2}{\varepsilon_{h}} \sum_{n=1}^{N}f(\mathbf{r}-\mathbf{r}_{n})%
\{ d\varepsilon_{n} \widetilde{E}(\mathbf{r}_n, \omega) -  \notag \\ %
\frac {\delta\varepsilon_{n}}{4} \left[ \widetilde{E}(\mathbf{r}_{n}, \omega+\Omega_{n}) +
 \widetilde{E}(\mathbf{r}_{n}, \omega-\Omega_{n}) \right] \} = -i\omega \frac{4\pi}{c^2}\widetilde{j}(\mathbf{r}, \omega), %
\label{osc6}
\end{eqnarray}
where
\begin{equation}
k(\omega)\equiv \frac{\omega }{c}\sqrt{\varepsilon _{h}}= \frac{2\pi }{\lambda },  \:%
d\varepsilon_{n} \equiv \varepsilon_{n}^{0}-\varepsilon_{h}+\delta \varepsilon_{n}/2, \: k(\omega)L_{n} \ll 1.
\label{osc6a}
\end{equation}
Here $k(\omega)$ is a wave number in the host medium, and $L_{n}$ is the characteristic size of the $n$-th scatterer (small particle).

Note that the Eq. (\ref{osc6}) is approximate one, and it is correct only when the small
scatterers are considered (so the condition $kL_{n}\ll 1$ should be satisfied).  The Eq. (\ref{osc6}) shows that we need to find three fields inside each particle ($\widetilde{E}(\mathbf{r}_{n}, \omega)$ and $\widetilde{E}(\mathbf{r}_{n}, \omega \pm \Omega_{n})$), and that the spectral parts of the fields become cross connected.

The solution of the Eq. (\ref{osc6}) is found as the sum of the incident $\widetilde{E}_{in}$ and the scattered $\widetilde{E}_{sc}$ fields
\begin{equation}
\widetilde{E}(\mathbf{r},\omega) = \widetilde{E}_{in}(\mathbf{r},\omega) + \widetilde{E}_{sc}(\mathbf{r},\omega),  %
\label{osc7}
\end{equation}
where the incident field $\widetilde{E}_{in}(\mathbf{r},\omega)$ is
\begin{equation}
\widetilde{E}_{in}(\mathbf{r},\omega) \equiv \frac{4 \pi i \omega}{c^2} %
\int_{-\infty }^{\infty }\frac{\widetilde{j}(\mathbf{q},\omega)%
e^{i\mathbf{q\cdot r}}}{q^{2}-k^{2}}d\mathbf{q},\;%
\widetilde{j}(\mathbf{q, \omega})\equiv \frac{1}{8\pi ^{3}}\int_{-\infty }^{\infty%
}j(\mathbf{r})e^{-i\mathbf{q\cdot r}}d\mathbf{r}, %
\label{osc8}
\end{equation}
and the field scattered by all $N$ particles $\widetilde{E}_{sc}$ is
\begin{eqnarray}
\widetilde{E}_{sc}(\mathbf{r},\omega)  \equiv  \sum_{n=1}^{N} \Phi _{n}(\mathbf{r}, \omega) %
 \{ d\varepsilon_{n} \widetilde{E}(\mathbf{r}_{n}, \omega) -    \label{osc9} \\%
 \frac {\delta \varepsilon_{n}}{4} \left[ \widetilde{E}(\mathbf{r}_{n}, \omega+\Omega_{n}) + %
 \widetilde{E}(\mathbf{r}_n, \omega-\Omega_{n}) \right] \}. \notag%
\end{eqnarray}
Here the function $\Phi_{n}$ is defined as
\begin{equation}
\Phi _{n}(\mathbf{r}, \omega)\equiv \frac{\omega^2}{c^2} \int_{-\infty }^{\infty }\frac{\widetilde{f}_{n}(%
\mathbf{q})e^{i\mathbf{q\cdot (r-r}_{n})}}{q^{2}-k^{2}}d\mathbf{q},
\label{osc10}
\end{equation}
where $\widetilde{f}_{n}(\mathbf{q})$ is the Fourier transform of the function $f_{n}(\mathbf{r})$ and it has the following form
\begin{equation}
\widetilde{f}_{n}(\mathbf{q})\equiv \frac{1}{8\pi ^{3}}\int_{-\infty }^{\infty %
}f_{n}(\mathbf{r})e^{-i\mathbf{q\cdot r}}d\mathbf{r}.  %
\label{osc10aa}
\end{equation}

The scattered field (\ref{osc9}) can be presented in more explicit form for the observer positioned outside of the cluster of the particles when $\mathbf{r} \neq \mathbf{r}_{n}$. In this case, the scattered field has the simplified form
\begin{eqnarray}
\widetilde{E}_{sc}(\mathbf{r},\omega)  =  \frac {\omega^{2}}{4 \pi c^{2}} \sum_{n=1}^{N} \frac{V_{n} e^{ikR_{n}}}{R_{n}} %
 \{ d\varepsilon_{n} \widetilde{E}(\mathbf{r}_{n}, \omega) -    \label{osc10d} \\%
 \frac {\delta \varepsilon_{n}}{4} \left[ \widetilde{E}(\mathbf{r}_{n}, \omega+\Omega_{n}) + %
 \widetilde{E}(\mathbf{r}_n, \omega-\Omega_{n}) \right] \}, \notag%
\end{eqnarray}
where distance from the observer to the $n$-th particle is
\begin{equation}
R_{n} \equiv |\mathbf{R}_{n}|, \; \mathbf{R}_{n} \equiv  \mathbf{r} - \mathbf{r}_{n} \neq 0.%
\label{osc10a30}
\end{equation}

Note, that the expressions (\ref{osc7})-(\ref{osc9}) form the complete solution of the scattering problem,
and the expressions for the scattered fields (\ref{osc9}) and (\ref{osc10d}) are the main results of this section. The formulae show that the scattered field $\widetilde{E}_{sc}(\mathbf{r},\omega)$ depends on different fields inside the particle: the field $\widetilde{E}(\mathbf{r},\omega)$ at the frequency $\omega$, and another two fields $\widetilde{E}(\mathbf{r},\omega \pm \Omega_{n})$ at the frequencies $\omega \pm \Omega_{n}$.

The expression (\ref{osc10d}) suggests that the scattered field can be zero at some frequencies when the following equation is satisfied for each particle
\begin{equation}
d\varepsilon_{n} \widetilde{E}(\mathbf{r}_{n}, \omega) =%
\frac {\delta \varepsilon_{n}}{4} \left[ \widetilde{E}(\mathbf{r}_{n}, \omega+\Omega_{n}) + %
 \widetilde{E}(\mathbf{r}_n, \omega-\Omega_{n}) \right] .
\label{osc10a40}
\end{equation}

As formulae (\ref{osc7}), (\ref{osc9}),  and (\ref{osc10d}) suggest, in order to calculate the scattered
field $\widetilde{E}_{sc}(\mathbf{r},\omega)$, we need to know not only the
fields $\widetilde{E}(\mathbf{r}_{n}, \omega)$ inside the $n$-th particle at frequency $\omega$, but
also the fields $\widetilde{E}(\mathbf{r}_{n}, \omega \pm \Omega_{n})$ at the frequencies $\omega\pm\Omega_{n}$.
To find these fields it will be required to solve the system of linear
equations with respect to the unknown fields inside the particles $\widetilde{E}(\mathbf{r}_{n}, \omega)$, and
$\widetilde{E}(\mathbf{r}_{n}, \omega \pm \Omega_{n})$. Note that, in principe, the system of the linear equations is infinite, because in order to calculate the fields $\widetilde{E}(\mathbf{r}_{n}, \omega \pm \Omega_{n})$ one needs
to know also the fields $\widetilde{E}(\mathbf{r}_{n}, \omega \pm 2\Omega_{n})$, and so on. This kind of chain connection also appears in the band structure calculations (see, for example \cite{Sakoda} and \cite{Kittel}),
and in other physical phenomena involving solution of the second order differential equations \cite{Patankar}.

To avoid this 'infinite chain problem' we have to truncate the system of the equations with respect to the number
of the aliasing frequencies. We define the aliasing frequencies as ones with values $\pm m\Omega_{n}$ where $1 \leq m \leq M$, and $M$ is the maximal number of the aliasing frequencies we have to take into account to ensure the minimal allowed error.

Another important reason to truncate the system of the equations, is that the LPM
condition $(kL_{n}\ll 1)$ must be valid at used frequencies $\omega \pm M\Omega_{n}$, resulting in the upper limit for the used frequency $\omega_{max}\sim 0.1c/ L_{n} \sqrt{\varepsilon _{h}} $.

For completeness, we present the system of equations for the fields $\widetilde{E}(\mathbf{r}_{n}, \omega)$
and $\widetilde{E}(\mathbf{r}_{n}, \omega \pm \Omega_{n})$ inside the particles located at the points $\mathbf{r}_{n}$
\begin{eqnarray}
\widetilde{E}(\mathbf{r}_{j},\omega) = \widetilde{E}_{in}(\mathbf{r}_{j},\omega) + \sum_{n=1}^{N}%
\{\alpha_{jn}(\omega) \widetilde{E}(\mathbf{r}_{n},\omega) \notag \\ %
- \beta_{jn}(\omega) \left[ \widetilde{E}(\mathbf{r}_{n},\omega + \Omega_{n}) + %
\widetilde{E}(\mathbf{r}_{n},\omega - \Omega_{n}) \right] \}, \; (1 \leq j \leq N)%
\label{osc11}
\end{eqnarray}
where the coefficients $\alpha_{jn}$ and $\beta_{jn}$ are
\begin{eqnarray}
\alpha_{jn}(\omega) \equiv d\varepsilon_{n} \frac {\omega^2}{c^2} \int_{-\infty }^{\infty} %
\frac {\widetilde{f}_{n}(\mathbf{q})}{q^2-k^2}e^{i\mathbf{q} \cdot \mathbf{R}_{jn}}d\mathbf{q}, %
\;  \beta_{jn}(\omega) \equiv \frac {\delta\varepsilon_{n}}{4} \frac {\alpha_{jn}(\omega)}{d\varepsilon_{n}}.  %
\label{osc12} %
\end{eqnarray}
Here we used the definition $\mathbf{R}_{jn} \equiv \mathbf{r}_{j}-\mathbf{r}_{n} $, and the condition that permittivity of each scattering particle is different and modified with its own frequency $\Omega_{n}$.

The expression (\ref{osc11}) for the fields inside the scatterers can be used to give simple estimation of the effect
of the oscillation. When $\Omega_{n} \ll \omega$, the formula (\ref{osc11}) can be rewritten in the simplified form
\begin{eqnarray}
\widetilde{E}(\mathbf{r}_{j},\omega) = \widetilde{E}_{in}(\mathbf{r}_{j},\omega) + \sum_{n=1}^{N}%
\{  \alpha_{jn}^{0}(\omega) \widetilde{E}(\mathbf{r}_{n},\omega) \notag \\ %
- 2 \beta_{jn}(\omega) \left[ \frac {\partial^{2} \widetilde{E}(\mathbf{r}_{n},\omega)}{2!\partial \omega^{2}} \Omega_{n}^{2} +  \frac {\partial^{4} \widetilde{E}(\mathbf{r}_{n},\omega)}{4!\partial \omega^{4}} \Omega_{n}^{4} + ...  \right]  \}, \; (1 \leq j \leq N).%
\label{osc11a1}
\end{eqnarray}
where
\begin{eqnarray}
\alpha_{jn}^{0}(\omega)  \equiv  \alpha_{jn}(\omega) \vert_{ \delta \varepsilon_{n} = 0 }. %
\label{osc11a5}
\end{eqnarray}
The expression (\ref{osc11a1}) suggests that the impact of the permittivity oscillations grows nonlinearly with $\Omega_{n}$ and it is more pronounced near a resonance where $\frac {\partial \widetilde{E}^{2}}{\partial \omega^{2}} \Omega_{n}^{2}$ may exceed the field $\widetilde{E}(\mathbf{r}_{n},\omega)$.

Moreover, the expression (\ref{osc11a1}) also suggests that the total scattered field will be zero when the field $\widetilde{E}(\mathbf{r}_{n},\omega)$ inside each $n$-th particle satisfies the following equation
\begin{eqnarray}
\alpha_{jn}^{0}(\omega) \widetilde{E}(\mathbf{r}_{n},\omega) = 2 \beta_{jn}(\omega) \left[ \frac {\partial^{2} \widetilde{E}(\mathbf{r}_{n},\omega)}{2!\partial \omega^{2}} \Omega_{n}^{2} +  \frac {\partial^{4} \widetilde{E}(\mathbf{r}_{n},\omega)}{4!\partial \omega^{4}} \Omega_{n}^{4} + ...  \right]. %
\label{osc11a10}
\end{eqnarray}
The equation (\ref{osc11a10}) is linear differential equation with respect to the field $\widetilde{E}(\mathbf{r}_{n},\omega)$, and its non trivial solutions can be found in the following form
\begin{eqnarray}
\widetilde{E}(\mathbf{r}_{n},\omega) =A\exp^{ i (a + ib) \omega}, \: (b>0).%
\label{osc11a15}
\end{eqnarray}
Finally, it is worth to compare the obtained results with the static case when there is no modification of
the refractive index, i.e. when $\delta \varepsilon_{n} \to 0$, or $\Omega_{n} \to 0$. In these cases, for
the scattered fields (\ref{osc10d}) and for the fields inside the particles (\ref{osc11}) we get respectively
\begin{eqnarray}
\widetilde{E}_{sc}(\mathbf{r},\omega)  = \sum_{n=1}^{N} (\varepsilon _{n}^{0}-\varepsilon_{h}) %
\Phi _{n}(\mathbf{r}, \omega) \widetilde{E}(\mathbf{r}_{n}, \omega), \label{osc20a} \\
 \widetilde{E}(\mathbf{r}_{j},\omega) = \widetilde{E}_{in}(\mathbf{r}_{j},\omega) + \sum_{n=1}^{N}%
\alpha_{jn}(\omega) \widetilde{E}(\mathbf{r}_{n},\omega). \label{osc20b} %
\end{eqnarray}
Note that the expressions (\ref{osc20a}) and (\ref{osc20b}) reproduce the relevant formulae for the
scattering by cluster of small particles (see, for example \cite{VP}).

It is worth to mention that by using the similar approach, it seems feasible to study the scattering by the
particles with permittivities having more general time-dependent form. Main condition here
is that permittivity of the particle should be expanded into finite Fourier series as
\begin{equation}
\varepsilon(t) = \varepsilon^{0} + \delta\varepsilon \sum \limits_{p=-P}^P{c_{p} e^{i\Omega_{p} t}},
\label{osc10a}
\end{equation}
where $c_{p}$ and $\Omega_{p}$ are the Fourier coefficients and frequencies for the permittivity $\varepsilon(t)$ of the particle.

\begin{figure}[t]
  \centering
  \begin{minipage}[b]{0.49\textwidth}
    \includegraphics[width=\textwidth]
    {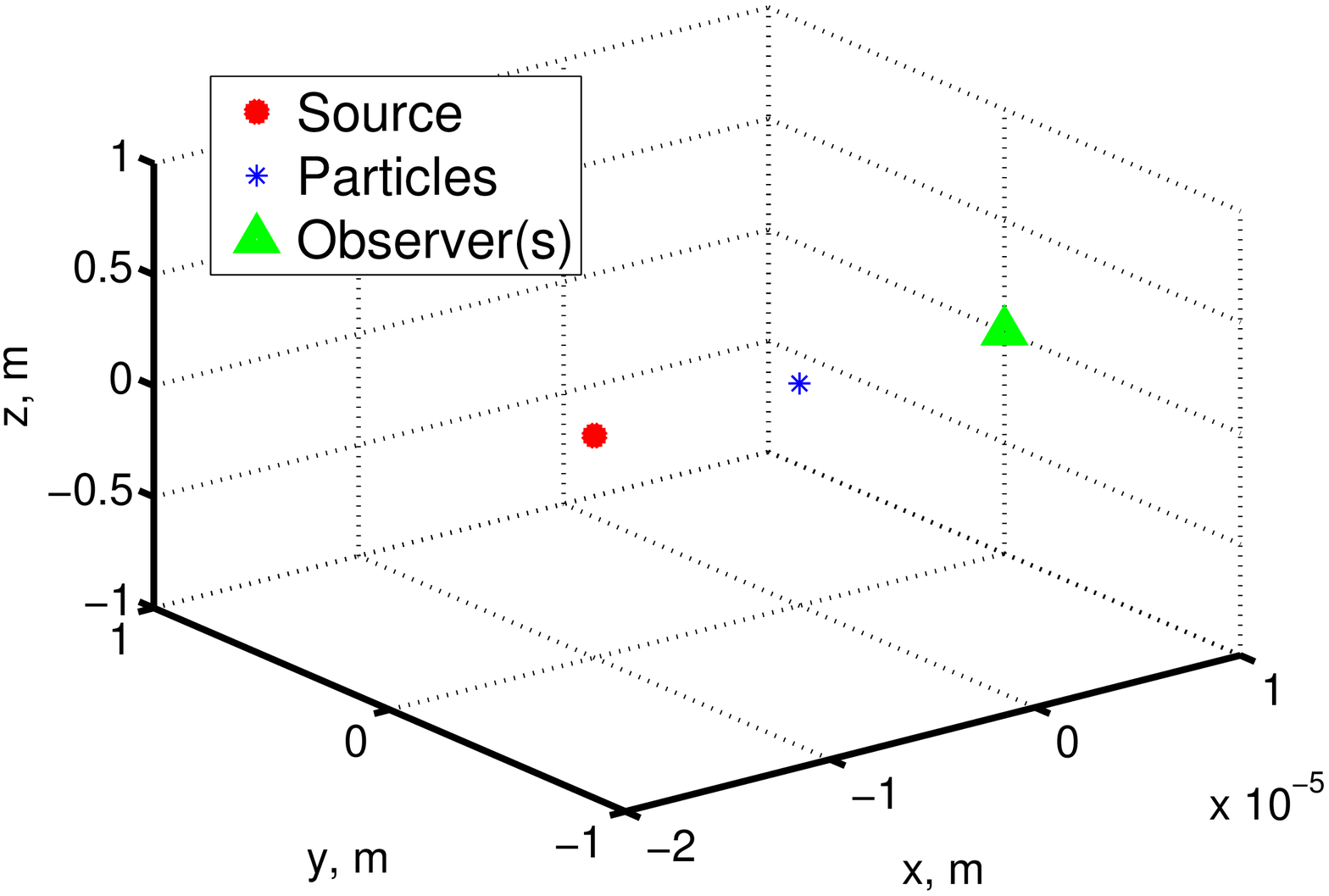}
  \end{minipage}
  \hfill
  \begin{minipage}[b]{0.49\textwidth}
    \includegraphics[width=\textwidth]
    {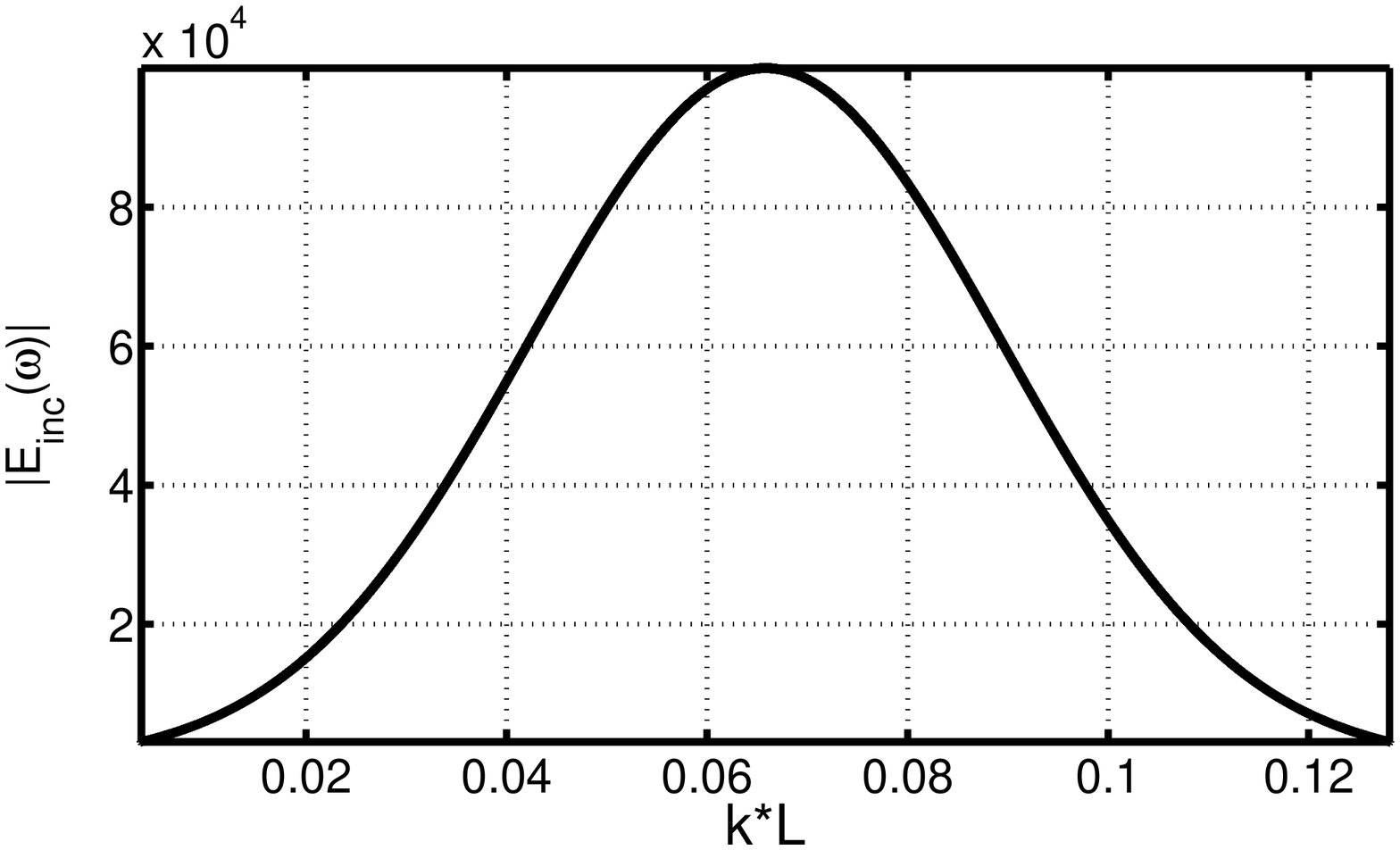}
  \end{minipage}
  \caption{The geometry of the scattering (positions of source, object, and observer) and the Gaussian spectrum of the incident field used in the numerical modeling.}
  \label{fig1}
\end{figure}

\section{Analysis for single particle with oscillating refractive index}
\subsection{The fields inside the particle}
In this subsection we apply the formulae obtained in the previous section for the scattering by small scatterer with
the oscillating refractive index. For the definiteness we assume that $\varepsilon_{n}^{0}-\varepsilon_{h} > \delta \varepsilon_{n}/2$.
\begin{figure}[t]
  \centering
    \includegraphics[width=\textwidth]
    {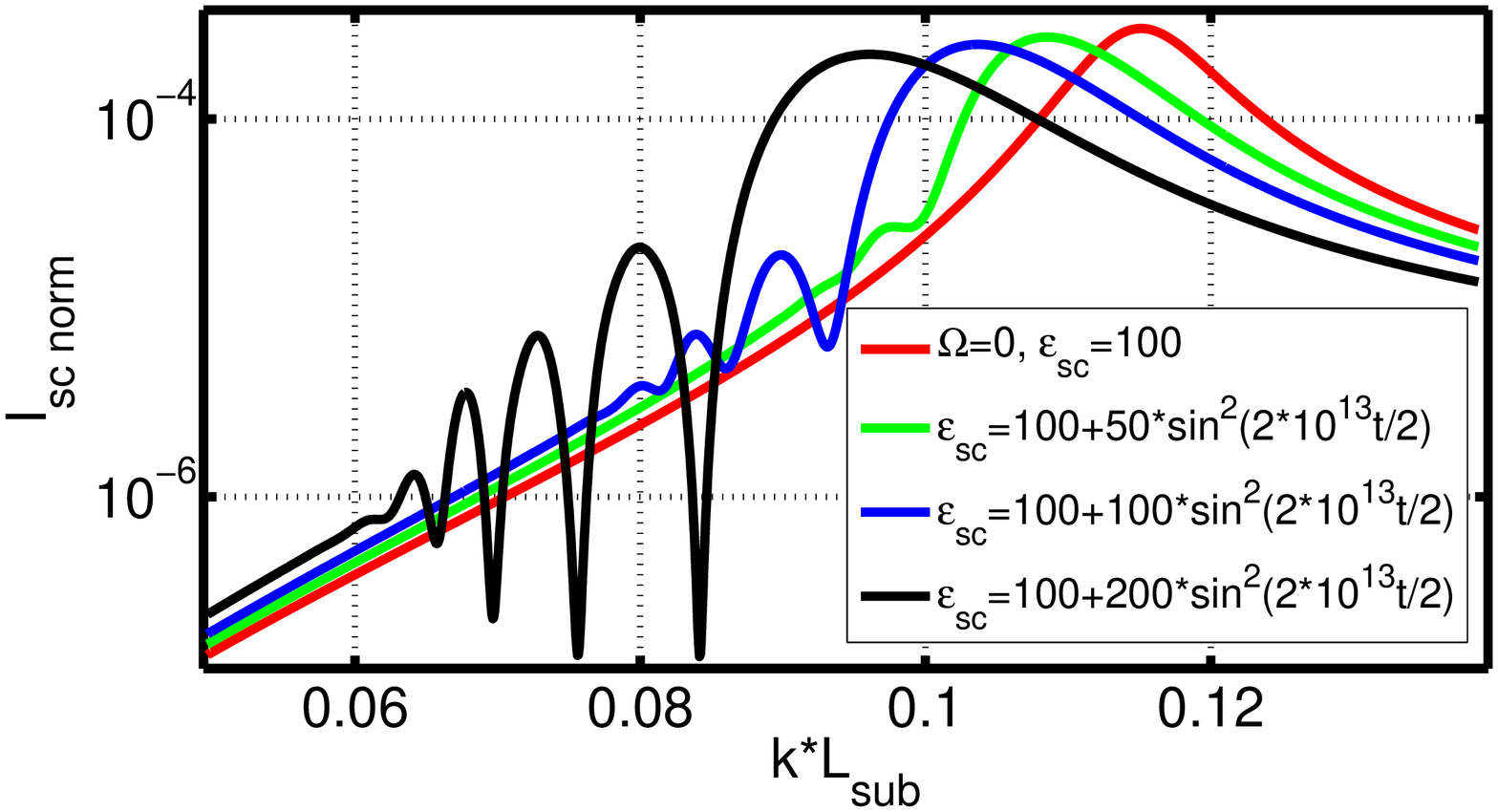}
  \hfill
  \caption{The normalized intensity $I_{sc \: norm}$ of the field scattered by the small cube with oscillating refractive index versus normalized frequency $kL$. The cube is positioned at the origin of coordinates ($\mathbf{r}_{1}=0$), and the width of the cube is $2L=40$ nm. The scattered fields are calculated by using sufficient number of aliasing frequencies ($M=32$) to guarantee the convergence of the results. The permittivity of the cube and the host medium respectively are $ \varepsilon _{n}=100+\delta\varepsilon\sin^{2}(2*10^{13}t/2)$ and $\varepsilon _{h}=1$, where
  $ \delta\varepsilon$=0, 50, 100, and 200. When the oscillation amplitude $\delta \varepsilon_{n}$ increases, several differences are seen in comparison with the reference case (cube with constant permittivity, solid curve): the shift of the resonances towards longer wavelengths, the emergence of multiple resonances (near $kL=0.08$, for example), and the appearing of the deep dives in the scattering spectrum (near $kL=0.075$ and $kL=0.085$, for example). }
  \label{fig2}
\end{figure}

Let us take a closer look at the fields inside the scattering particle. Suppose that the particle
is located at the point $\mathbf{r}_{1}$, and its permittivity oscillates with the frequency $\Omega$.
To find the fields inside the particle we have to solve the truncated system of $2M+1$ linear equations with respect to the unknown fields $\widetilde{E}(\mathbf{r}_{1},\omega \pm m\Omega)$ with $0 \leq m \leq M$. The matrix of the coefficients is tridiagonal, and the matrix dimension is $(2M+1) \times (2M+1)$.

\begin{figure}[t]
  \centering
    \includegraphics[width=\textwidth]
    {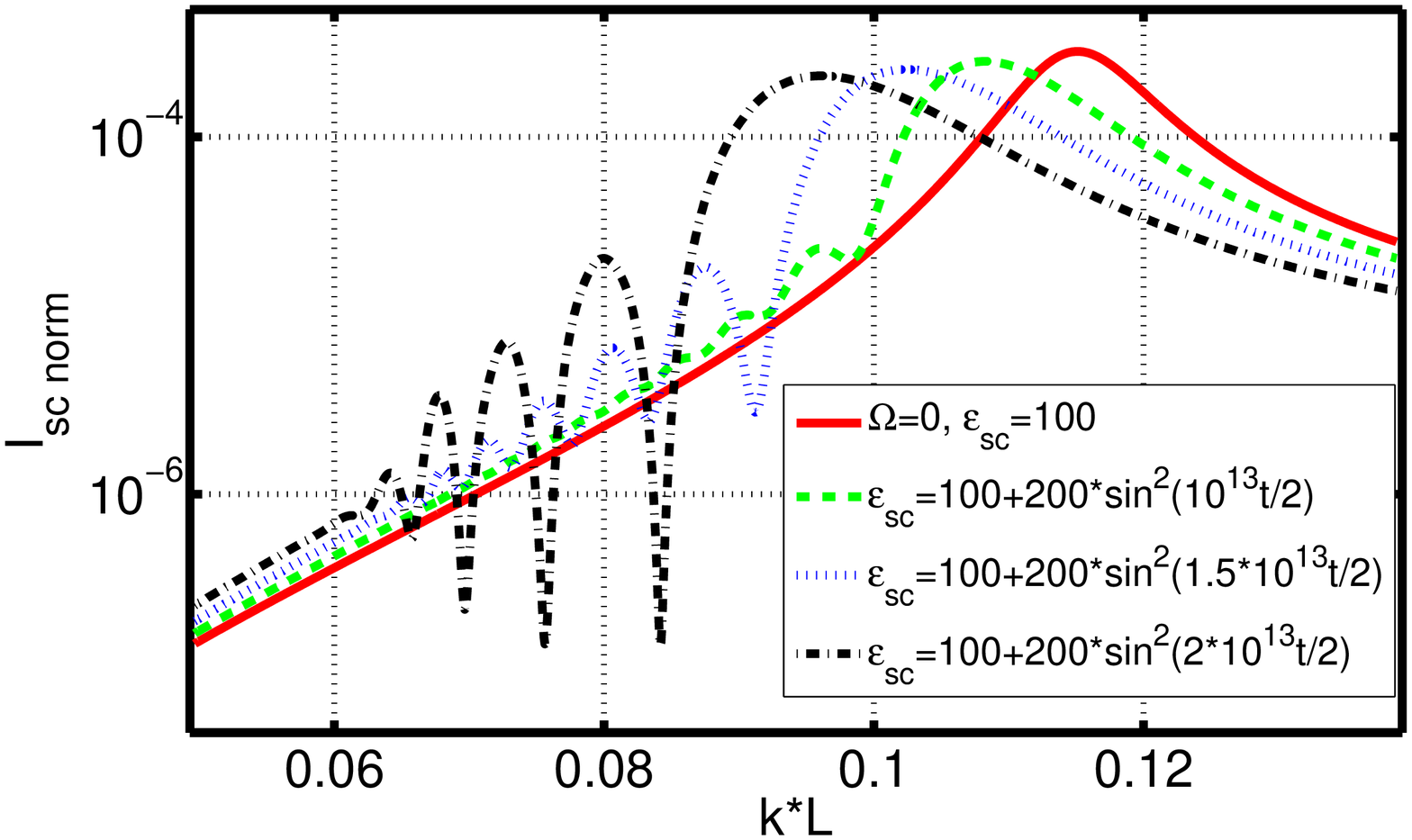}
  \hfill
  \caption{The normalized intensity $I_{sc \: norm}$ of the field scattered by the small cube with oscillating refractive index versus normalized frequency $kL$. The cube is positioned at the origin of coordinates ($\mathbf{r}_{1}=0$), and the width of the cube is $2L=40$ nm. The scattered fields are calculated by using sufficient number of aliasing frequencies ($M= 64$) to guarantee the convergence of the results. The permittivity of the cube and the host medium respectively are $ \varepsilon _{n}=100+200\sin^{2}(\Omega t/2)$ and $\varepsilon _{h}=1$, where
  $ \Omega$=0, 10, 15, and 20 THz. When the oscillation frequency $\Omega$ increases, several differences are seen in comparison with the reference case (cube with constant permittivity, solid curve): the shift of the resonances towards longer wavelengths, the emergence of multiple resonances (near $kL=0.08$, for example), and the appearing of the deep dives in the scattering spectrum (near $kL=0.075$ and $kL=0.085$), for example.}
  \label{fig2a}
\end{figure}

 To get explicit results for analysis, we limit the number of the aliasing frequencies by $M=2$, and the
 solutions for the fields inside the particle becomes
\begin{equation}
\widetilde{E}(\mathbf{r}_{1}, \omega)  = \frac { \widetilde{E}_{inc}(\mathbf{r}_{1},\omega) -
\beta_{11}(\omega) \left(  \frac{A_{+}}{\gamma_{+}} + \frac{A_{-}}{\gamma_{-}} \right) } %
{(1-\alpha_{11}(\omega))D(\omega, \Omega)} ,  \label{osc22}
\end{equation}
and
\begin{equation}
\widetilde{E}(\mathbf{r}_{1}, \omega + \Omega) + \widetilde{E}(\mathbf{r}_{1}, \omega - \Omega) = %
\frac {A_{+}}{\gamma_{+}} + \frac {A_{-}}{\gamma_{-}}- %
\widetilde{E}(\mathbf{r}_{1}, \omega) \left( \frac{\xi_{1\Omega}}{\gamma_{+}} + %
\frac{\xi_{-1\Omega}}{\gamma_{-}} \right),
\label{osc24} %
\end{equation}
where the denominator $D(\omega, \Omega)$ is defined as
\begin{equation}
D(\omega, \Omega) \equiv  1 - \xi_{0\Omega} \left( \frac {\xi_{1\Omega}} {\gamma_{+}} + %
\frac {\xi_{-1\Omega}}  {\gamma_{-} }\right).   \label{osc22a} %
\end{equation}
The coefficients $A_{\pm}$ and $\gamma_{\pm}$ are
\begin{equation}
A_{\pm} \equiv F_{\pm 1\Omega} -\xi_{\pm 1\Omega}F_{\pm 2\Omega}, \: \:%
\gamma_{\pm} \equiv 1-\xi_{\pm 1\Omega}\xi_{\pm 2\Omega},  %
\label{osc25} %
\end{equation}
and the coefficients $F_{\pm m \Omega}$ and $\xi_{\pm m \Omega}$ are
\begin{eqnarray}
 F_{\pm m \Omega} \equiv \frac { \widetilde{E}_{inc}(\mathbf{r}_{1}, \omega \pm m \Omega) } {1-\alpha_{11}(\omega \pm m \Omega)}, %
  \label{osc25g} \\ %
 \xi_{\pm m \Omega} \equiv \frac {\beta_{11}(\omega \pm m \Omega) } {1-\alpha_{11}(\omega \pm m \Omega)}, %
  (0\leq m \leq 2).%
\label{osc26} %
\end{eqnarray}
The formulae (\ref{osc22}), and (\ref{osc24}) show that the field inside the particle at frequency $\omega$ is a function of the incident fields at optical frequencies $\omega\pm m\Omega$.
As the result, the fields inside scatterer may have resonances not only when $Re(\alpha_{11}(\omega))=1$ (as in stationary case), but also when $Re(\alpha_{11}(\omega \pm m\Omega))=1$, or $Re(D(\omega), \Omega)=0$.

We note, that fields at $M=1$ can be obtained from the solutions for $M=2$ by setting $F_{\pm 2\Omega}=0$ and
 $\xi_{\pm 2\Omega}=0$. We will use this in the following discussion.

The solutions (\ref{osc22}), and (\ref{osc24}) suggest also that we can limit the number of the aliasing frequencies when $|\xi_{(M-1) \Omega}\xi_{ M \Omega}| \ll 1$.  To clarify this, we consider the
terms $|\xi_{M \Omega}|^{2} \approx |\xi_{(M-1) \Omega}\xi_{ M \Omega}|$ for small sphere
 in detail when $M \gg 1$, and $\varepsilon_{h} \sim 1$. Taking into account that for the
small sphere the coefficient $\alpha_{11}$ is
\begin{equation}
\alpha_{11}(\omega)=\frac{k^{2}L^{2}d\varepsilon}{2\varepsilon_{h}} \left(1+2ikL/3 \right),
\label{osc27} %
\end{equation}
and by using the Eq. (\ref{osc26}), we estimate that far from a resonance
\begin{equation}
|\xi_{ M \Omega}|^{2} \approx k_{\Omega}^{4} L^{4} \frac{ |\delta \varepsilon|^{2} }{2^{6}}, \:%
k_{\Omega}^{2} \equiv \frac {(\omega+ M \Omega)^{2}}{c^{2}}, \left( M \gg 1, \varepsilon_{h} \sim 1 \right).%
\label{osc28} %
\end{equation}
The Eq. (\ref{osc28}) shows that far from a resonance, the condition $|\xi_{ M \Omega}|^{2} \ll 1$ is satisfied when
$k_{\Omega}^{2} L^{2} |\delta \varepsilon|  \leq 1$.
While near the resonance, when $1=Re(\alpha_{11}(\omega))$, we have to take into account nonzero imaginary part of the coefficient $\alpha_{11}$, and we get
\begin{equation}
|\xi_{ M \Omega}|^{2} \approx  %
\frac {|\delta \varepsilon|^{2} } {7\left[ k_{\Omega}LRe(d\varepsilon)+\frac{3}{2}Im(d\varepsilon) \right]^2}. %
\label{osc29} %
\end{equation}
The expression (\ref{osc29}) shows that despite the strict LPM condition $k_{\Omega}L \ll 1$,
it is relatively easy to satisfy the condition $|\xi_{ M \Omega}|^{2} \ll 1$ for single particle when
\begin{equation}
\frac {|\delta \varepsilon| }{\left[ k_{\Omega}LRe(d\varepsilon)+\frac{3}{2}Im(d\varepsilon) \right]} \leq 1. %
\label{osc29a} %
\end{equation}
When the inequation (\ref{osc29a}) is satisfied, one can use very limited number of the aliasing frequencies $M$ while maintaining broad parameter space. When the inequation (\ref{osc29a}) is not satisfied, the large number of the aliasing frequencies $M$ may be needed to obtain correct results. For example, when $\varepsilon_{1}^{0}=100$ and $\delta \varepsilon_{1}=10$, the inequality (\ref{osc29}) is satisfied, and the coefficient $|\xi_{ M \Omega}|^{2} \approx 0.13$. So, in this case, the number of aliasing frequencies $M$ can be limited to only few.

\begin{figure}[t]
  \centering
    \includegraphics[width=\textwidth]
    {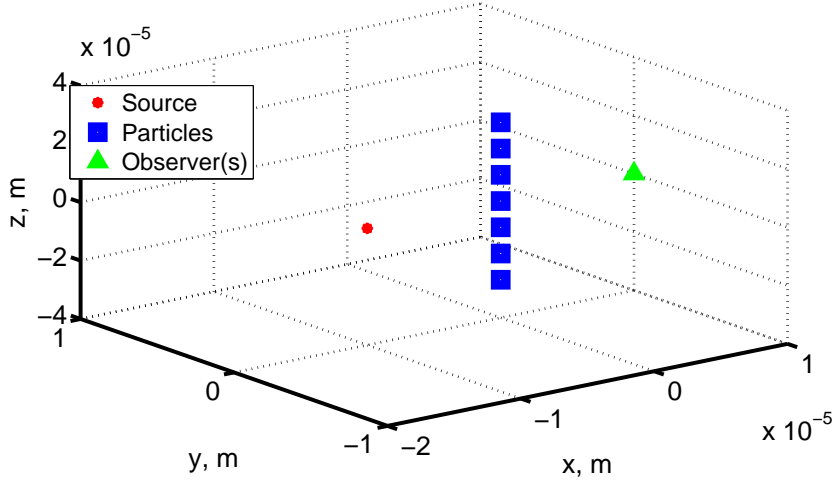}
  \hfill
  \caption{The geometry of the scattering by 7 small cubes aligned along $z$ direction. The width of each cube is
  $2L=40$ nm and the distance between adjacent cubes is $d=9$ $\mu$m.}
  \label{fig3}
\end{figure}

\subsection{The resonance frequency and the resonance width}
It should be noted that the resonance frequency is important indicator of the influence of the permittivity
oscillating on the scattering. Below we analyse the resonance frequency for the field scattered
by small sphere and small cube. The expression $(\ref{osc22})$ gives the field inside the small particle at optical
frequency $\omega$, when the number of the aliasing frequencies is $M=2$.
Even this expression is too complex to study analytically, so to demonstrate
influence of the oscillations, we study the field $E(\mathbf{r}_{1}, \omega)$ at $M=1$, and
neglect by the terms $\xi_{0 \Omega}\xi_{\pm 1\Omega}$ in the denominator. Under these conditions,
the field inside the particle is
\begin{eqnarray}
\widetilde{E}(\mathbf{r}_{1}, \omega)  \approx  F_{0 \Omega} - \xi_{0 \Omega} %
\left(  F_{1\Omega} + F_{-1\Omega} \right)  = \notag \\%
\frac { \widetilde{E}_{inc}(\mathbf{r}_{1}, \omega)}{1-\alpha_{11}(\omega)} - %
\frac {\beta_{11}(\omega)}{1-\alpha_{11}(\omega)} %
\left[  \frac{ \widetilde{E}_{inc}(\mathbf{r}_{1},\omega+\Omega)}{1-\alpha_{11}(\omega+\Omega)} + %
\frac{ \widetilde{E}_{inc}(\mathbf{r}_{1},\omega-\Omega)}{1-\alpha_{11}(\omega-\Omega)}  \right]. %
\label{osc30} %
\end{eqnarray}
The approximated field (\ref{osc30}) has several maxima at resonance frequencies which are solutions of the equations $\alpha_{11}(\omega \pm m\Omega)=1$, with $m=0,1$.
When $m=0$, the corresponding resonance frequency is
\begin{equation}
\omega_{r0} = \left\{
\begin{array}{rl}
 \frac{\sqrt{2}c}{L} \frac {1} {\sqrt{\varepsilon_{n}^{0}-\varepsilon_{h}+\delta \varepsilon_{n}/2}},
 & \text{for sphere}\\
\frac{\sqrt{\pi}c}{1.54L} \frac{1}{\sqrt{\varepsilon_{n}^{0}-\varepsilon_{h}+\delta \varepsilon_{n}/2}},
& \text{for cube }
\end{array} \right.
\label{osc32}
\end{equation}
and when $m=1$, the resonance frequencies are
\begin{equation}
\omega_{r1} = \omega_{r0} \pm \Omega.  \label{osc32a} %
\end{equation}
The resonance frequency (\ref{osc32}) is a new result suggesting that $\omega_{r0}$ is a function of the static and dynamic parts of the refractive index of the particle, and it also reproduces well known formula for the resonance frequency of the particle (sphere or cube) with constant permittivity $\varepsilon _{n}^{0}$  \cite{VP}. The expression (\ref{osc32}) suggests that the resonance frequency $\omega_{r0}$ decreases or increases when the permittivity variation amplitude $\delta \varepsilon_{n}$ grows or decreases respectively.

The resonant frequencies (\ref{osc32a}) are new results and they take into account the oscillations of the permittivity with the frequency $\Omega$. The Eq. (\ref{osc30}) suggests that these new resonances will be visible only when parameter $\beta_{11}(\omega) \sim 1$.

For the function $f(\omega)=1/ \left| \Psi(\omega) \right|$ the resonance width $\xi $ can be estimated by using the following expression
\begin{equation}
\xi  \approx  \frac { 2 \sqrt{3} \left| \operatorname{Im}\Psi(\omega_{r}) \right|}
 { \left|     \frac  {  \partial \operatorname{Re}\Psi(\omega )  }
           { \partial \omega}   \right|_{\omega=\omega_{r}} },
\label{osc35}
\end{equation}
and will use this expression for our analysis. By using the expression (\ref{osc35}) and the expression for the resonance frequency (\ref{osc32}) adapted for small sphere, we estimate the resonance width as
\begin{equation}
\xi  \approx  \frac { 4} { \sqrt{3} } \frac { c \sqrt{\varepsilon_{h}}} {  L d\varepsilon_{n} }.
\label{osc35a}
\end{equation}
The formula (\ref{osc35a}) suggests that the resonance width decreases with the growth of the optical contrast $d\varepsilon_{n}$ of the particle, and for the sphere of the radius of $20$ nm, for example, the resonance width is about $\xi \approx 300$ THz. If oscillating frequency is much smaller than the resonance width ($ \Omega \ll  \xi $), the resonances described by the expression (\ref{osc32a}) can not be distinguished when $M \sim 1$.

\begin{figure}[t]
  \centering
  \hfill
    \includegraphics[width=\textwidth]
    {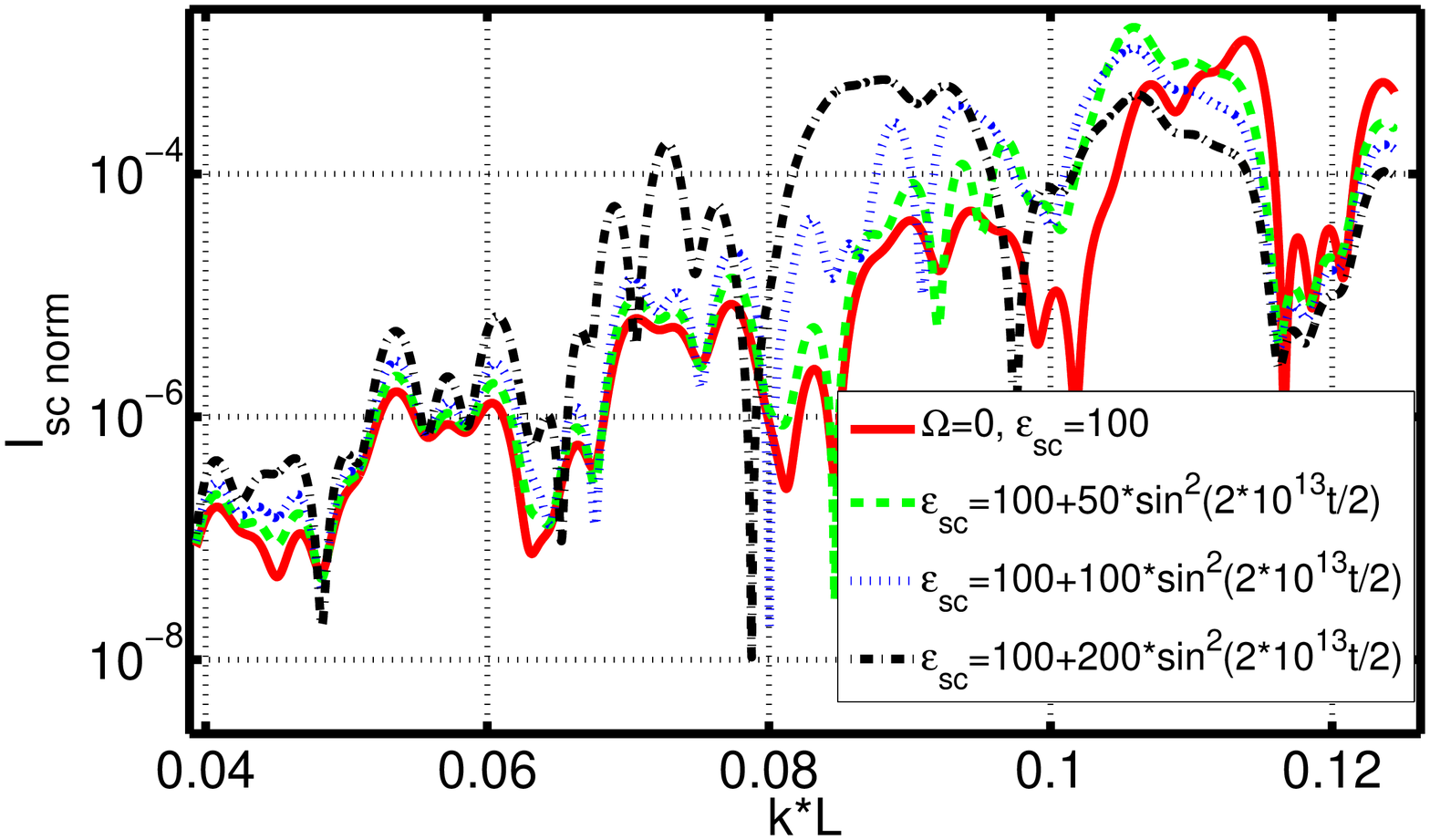}
  \caption{The normalized intensity $I_{sc \: norm}$ of the field scattered by 7 small cubes versus normalized frequency $kL$. The cubes are positioned along $z$ axis with period $d=9$ $\mu$m, and the width of each cube is
  $2L=40$ nm. The scattered fields are calculated at $M=32$ aliasing frequencies. The permittivities of the cubes and the host medium respectively are $\varepsilon _{n}=100+\delta \varepsilon \sin^{2}(2*10^{13} t/2)$ and $\varepsilon _{h}=1$, where $\delta \varepsilon=50, 100$, and $200$. When the oscillation amplitude $\delta \varepsilon_{n}$ increases, several differences are seen in comparison with the reference case (cubes with constant permittivity, solid curve): the emergence of multiple resonances (near $kL=0.06$, and $kL=0.072$, for example), and the appearing of the deep dives in the scattering spectrum (near $kL=0.08$, for example).}
  \label{fig3a}
\end{figure}

\section{Numerical modeling examples}

Theoretical investigations of the wave scattering can be done for few small particles, however, when the number of the particles is increased, the numerical modeling is the only practical way to get useful results.

In this section we numerically calculate the field scattered by one small cube and by cluster of small cubes with the oscillating refractive index. We will change two parameters: the oscillating frequency $\Omega$, and the amplitude of oscillations $\delta \varepsilon$.

\begin{figure}[t]
  \centering
  \hfill
    \includegraphics[width=\textwidth]
    {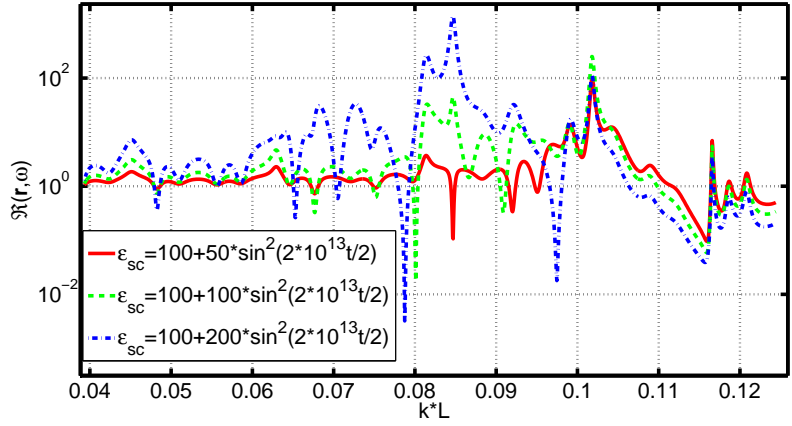}
  \caption{The ratio $\Re(\mathbf{r},\omega)$ of the intensity of the field scattered by system with oscillating permittivity and same system with constant permittivity. The system consists of 7 small cubes, and the ratio is plotted versus normalized frequency $kL$ for the observer positioned at the point $\mathbf{r}$. The cubes are positioned along $z$ axis with period $d=9$ $\mu$m, and the width of each cube is $2L=40$ nm. The scattered fields are calculated at $M=32$ aliasing frequencies. The permittivities of the cubes and the host medium respectively are $\varepsilon _{n}=100+\delta \varepsilon \sin^{2}(2*10^{13} t/2)$ and $\varepsilon _{h}=1$, where $\delta \varepsilon=50, 100$, and $200$. The ratio curves indicate the surging of resonances, and appearing of deeps in the scattering spectrum.}
  \label{fig3b}
\end{figure}

\subsection{Single particle example}

In the Fig.(\ref{fig1}) the geometry of the scattering and the spectral intensity of the incident field are shown. We put the particle between the point source and the observer in the center of coordinates $\mathbf{r}_1=0$, and the incident field has Gaussian profile centered in the middle of the used spectrum.

In the Fig.(\ref{fig2}) and Fig.(\ref{fig2a}) the normalized intensities of the scattered field
\begin{equation}
 I_{sc \: norm}(\mathbf{r},\omega) \equiv \frac{I_{sc}(\mathbf{r},\omega) }{I_{inc}(\mathbf{r},\omega)}=
 \frac{ \left| \widetilde{E}_{sc}(\mathbf{r},\omega)\right|^2 }{ \left| \widetilde{E}_{inc}(\mathbf{r},\omega)\right|^2 }
\label{osc37a}
\end{equation}
are presented for the small cube with a side length of $2L=40$ nm. In the Fig.(\ref{fig2}) the results are presented for the cube with permittivity $\varepsilon _{n}=100+\delta \varepsilon\sin^{2}(2*10^{13}t/2)$, where $ \delta\varepsilon$=0, 50, 100, and 200, and in the Fig.(\ref{fig2a}) the permittivity of the scattering cube was set as $\varepsilon _{n}=100+200\sin^{2}(\Omega t/2)$, where $ \Omega$=0, 10, 15, and 20 THz.

The obtained results suggest that increasing the modulation amplitude $\delta \varepsilon$ or frequency $\Omega$ we shift the resonance towards longer wavelengths, as tentatively predicted by the analytical formulae $(\ref{osc32})$ and $(\ref{osc32a})$. These results also suggest that at higher modulation amplitudes and frequencies, the scattering becomes much more complex: additional resonances grow, and significant deeps in the scattered intensity appear (up to 10 times in comparison with the reference intensity). The results also indicate that number of the aliasing frequencies $M$ needed to be taken into account increases with the amplitude of the permittivity modulations.

\subsection{Many particles example}
In this subsection we present results of the numerical calculation of the field scattered by object made of multiple small particles. We model the scattering by 7 identical cubes with periodically oscillating refractive index (the geometry of the scattering is shown in Fig. (\ref{fig3})). The side length of a cube is $2L=40$ nm, and the cubes are aligned along $z$ direction and located in $yz$ plane. The distance between adjacent cubes is set to $d=9$ $\mu$m to see multiple resonances in the scattered field.

In the Fig.(\ref{fig3a}) the normalized intensities of the scattered fields are presented. To see the impact of the oscillations more clearly, we also presented the ratio $\Re$ of the intensities of the scattered fields in the Fig.(\ref{fig3b}). The ratio $\Re$ is defined as
\begin{equation}
\Re(\mathbf{r},\omega) \equiv \frac{\left| \widetilde{E}_{sc}(\mathbf{r},\omega) \right|^{2}_{\Omega=0}} %
   {\left| \widetilde{E}_{sc}(\mathbf{r},\omega) \right|^{2}_{\Omega=2*10^{13}, \: \delta \varepsilon=50, \: 100, \: 200}}.
\label{osc40}
\end{equation}

As for the scattering by single particle, the results obtained here suggest that at higher modulation amplitudes $\delta \varepsilon$ and frequencies $\Omega$, the scattering becomes much more complex: existing resonances shift, additional resonances grow, and significant deeps (up to 3 orders of magnitude) in scattered intensity appear.

\section{Conclusions}

The fields scattered by the cluster of small particles with oscillating refractive index have been studied theoretically
by using the local perturbation method in scalar approximation. The resonance width, and the resonance frequencies of the field scattered by the small particles have been calculated, and it has been shown that they depend on the amplitude and the frequency of the refractive index oscillations.
The scattering by single small cube and by the cluster of the small cubes with oscillating refractive index has been numerically calculated. The condition for the convergence of the numerical modeling has been discussed.
The theoretical and the numerical results suggest that the oscillating refractive index significantly affects the scattered field: existing resonances shift, additional resonances emerge, and new deeps in the scattering spectrum appear.

\begin{equation*}
\end{equation*}

\textbf{Acknowledgments}

Many thanks to my wife Lucy for her understanding and patience, and to my mother Lyudmila for her unconditional support.

\end{document}